\documentclass{aastex}
\usepackage{spr-astr-addons}
\usepackage{url}

\RequirePackage{color}

\begin{document}

\title{On the accretion disc properties in eclipsing dwarf nova EM Cyg}
\slugcomment{}
\shorttitle{On the accretion disc properties in EM Cyg}
\shortauthors{Halevin et al.}

\author{A.V. Halevin}
\author{I.I. Solovieva} 
\affil{Department of Astronomy, Odessa National University, T.G.Shevchenko Park, Odessa, 65014, Ukraine}
\and
\author{P.A. Dubovski} \and \author{I. Kudzej}
\affil{Vyhorlat Observatory, Mierova 4, Humenne, The Slovak Republic}
\email{alex.halevin@gmail.com} 

\begin{abstract}
In this paper we analyzed the behavior of the unusual dwarf nova EM Cyg using the data obtained in April-October, 
2007 in Vyhorlat observatory (Slovak Republic) and in September, 2006 in Crimean Astrophysical Observatory (Ukraine). 
During our observations EM Cyg has shown outbursts in every 15-40 days. 
Because on the light curves of EM Cyg the partial eclipse of an accretion disc is observed we applied the eclipse mapping technique 
to reconstruct the temperature distribution in eclipsed parts of the disc. Calculations of the accretion rate in the system were made 
for the quiescent and the outburst states of activity for different distances.
\end{abstract}

\keywords{accretion, accretion discs - binaries: close - binaries: eclipsing - stars: dwarf novae - stars: individual:
EM Cygni - cataclysmic variables.}


\section{Introduction}

EM Cyg is an eclipsing dwarf nova which has relatively long orbital period ($P_{orb}= 6^h.98$).
Dwarf novae are evolved binary stars, where the Roche lobe-filling red dwarf component accretes matter onto 
the white dwarf component. Loosing angular momentum, matter forms an accretion disc. Such systems  
produce outbursts, caused by instabilities in the disc.  
 
EM Cyg was discovered by Hoffmeister in 1928. \citet{mk69} investigated instabilities 
on the eclipse light curves and noted, that they reflect instabilities in accretion disc. 
\citet{sti82} detected rapid oscillations on light curves with typical time scale of 14.6 and 16.5 seconds.

\citet{br77}, using long-term photographic observations 
found outburst cycle to be 24-25 days and noted the seasonal changes of the light curve and the outburst cycle.

Because EM Cyg shows standstills \citep{nor00}, it is classified as Z Cam type star. So EM Cyg is the only eclipsing dwarf nova among Z Cam type stars.

Using radial velocities, \citet{rob74} estimated 
masses to be 0.7 $M_{\odot}$ and 0.9 $M_{\odot}$ for the white dwarf and the red dwarf respectively. These parameters corresponded to 
the thermal-timescale mass transfer in the system, and were doubtful. 
\citet{sto81} found that the orbital inclination of the system is about $63^{\circ}$ and the mass relation close to results obtained by \citet{rob74}.

\citet{nor00} found that the spectrum of EM Cyg in the range 6230-6650 $\AA$ is contaminated by light from a K2-5V 
star (in addition to the K-type mass donor star). The K2-5V star contributes approximately 16 \% of the light from the system in that band and, if not 
taken into account, has a considerable effect upon radial velocity measurements of the mass donor star. The revised value of the mass ratio, 
combined with the orbital inclination $i=67^{\circ}$, leads to the masses of $0.99 M_{\odot}$ and $1.12 M_{\odot}$ for the mass donor 
and white dwarf respectively.  

Recent more precise measurements of the third star light by Welsh et al. (2007) showed that masses to be $M_{wd}=1.0 M_{\odot}$ and $M_{rd}=0.77 M_{\odot}$.

There are several too different estimates of the distance to EM Cyg.
From ellipsoidal variations in the infrared band, \citet{jks81} found secondary spectral type to be K2V and the distance to the system to be 320 pc.
\citet{bay81} using K magnitudes of \citet{jks81} determined the possible distance to be in the range 285-429 pc with
most probable value of about 352 pc.
But \citet{sti82}, using private communication of Stover suggests 400 pc in his calculations.

Taking into account the light of the third star, \citet{wel05} supposed the distance to be 450 - 500 pc.

Winter and Sion (2003) found 411 parsecs using the $M_v$ versus $P_{orb}$ relation \citep{war95}, but they determined only 210 parsecs when they fitted the
IUE spectra.

Latest paper of \citet{go09} give the distances ranged from 200 to 380 pc for different WD+disc models of UV spectra.

Model fits for ultraviolet data \citep{ws03, go09} give mass accretion rate estimates to be about $10^{-10} M_{\odot}/yr$. This value lies
below the upper limit ($2\cdot10^{-9} M_{\odot}/yr$), determined by \citet{cz08} from the O-C diagram analysis.

According to \citet{voi78}, there are no circular polarization above about 0.3\% in EM Cyg.

Despite low Galactic latitude (+4.28$^\circ$), the reddening estimate $E_{B-V}$ towards EM Cyg is only 0.03$\div$0.05 \citep{ver87, lad91, be94}.

\begin{figure}[t]
\includegraphics[width=84mm]{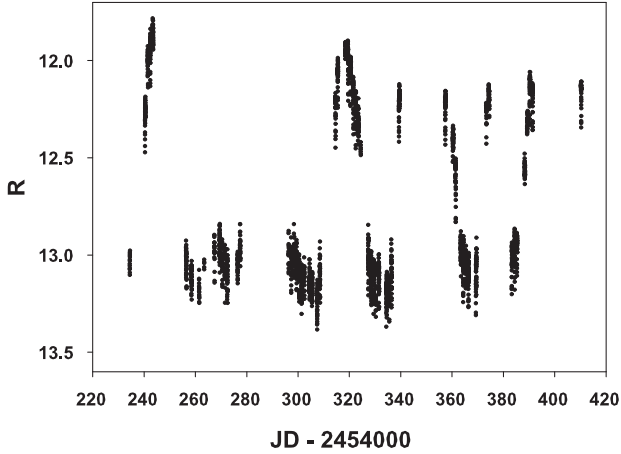}
\caption{Long term light curve of EM Cyg, obtained in 2007 in Vyhorlat observatory  (Humenne, Slovak Republic).}
\end{figure}

\section{Observations}

Our photometric observations of EM Cyg were obtained during September, 2006 using the AZT-11 telescope (d=1.25m, Cassegrain system) of the Crimean Astrophysical Observatory (Ukraine)
with the FLI CCD camera in R band with 15 sec exposure and using the Pupava telescope (d=280 mm, Newton system) 
with the MEADE DSI PRO CCD camera in R band with 30 sec exposure in Vyhorlat observatory (Slovak Republic) during May-November, 2007 
(Fig.1, see online database http://var.kozmos.sk). The 
quantity of observational runs are 8 for Crimean data and 61 for Slovakian data.

CCD images were processed with C-Munipack package \citep{hro98}, including flatfielding and debiasing. The procedure of the "artificial'' mean weighted star 
\citep{ab04, kim05} was used to calculate of the instrumental magnitudes of EM Cyg. Using a secondary photometric standard from \citet{mis96}, we calculated  
transformation coefficients for converting the magnitudes from the instrumental system of Pupava telescope: 

$m-m_0 = -0.027(\pm0.012) + 0.98(\pm0.024)\cdot(V_0-R_0)$

Moments of minima we determined using the method of "asymptotic parabola" \citep{am06} (Table 1).  
To achieve a higher accuracy with period determination, we used 30 moments of minima obtained by the previous investigators too 
\citep{mk69, mum74, mum75, mum80, bp84, cz08}.

The improved elements are
\begin{equation}
T = 2437882.86059(33) + 0.290909199(84)^d \cdot E
\end{equation}

\begin{table}[t]
 \centering
  \center
  \caption{28 heliocentric moments of minima of EM Cyg (HJD-240000). (1) - Crimean data, (2): Slovakian data}
  \begin{tabular}{ccccc}
  \hline
$53947.44411^1$&$54269.48212^2$&$54328.53527^2$\\
$53953.40727^1$&$54272.38822^2$&$54331.44782^2$\\
$53954.41905^1$&$54299.44092^2$&$54336.39064^2$\\
$53960.38641^1$&$54314.57070^2$&$54339.29649^2$\\
$53961.40854^1$&$54315.44400^2$&$54357.33454^2$\\
$54240.38778^2$&$54319.51621^2$&$54361.40679^2$\\
$54242.42310^2$&$54320.38851^2$&$54364.31680^2$\\
$54256.38635^2$&$54321.55141^2$&$54366.35136^2$\\
$54258.43000^2$&$54322.42526^2$&\\
$54267.44340^2$&$54327.36756^2$&\\        

\hline         
\end{tabular}  
\end{table}   

From Fig.1, one can see that the typical interval between outbursts is about 25 days. 
Duration of outburst is approximately equal to the time interval between them. 

Individual light curves show significant flickering which deforms the partial eclipse shape.

On the individual light curves the effect of the presence of a bright hot spot 
in accretion disc is presented. The hump before the eclipses is well visible, and the ascending branch of the eclipse light curve is longer, 
than the descending one. 

The strong flickering and variability with time-scales from several minutes to dozen percents of the orbital period are presented on the light curves. 

To investigate the typical patterns on the orbital light curves of EM Cyg and to diminish the flickering distortions,
all phase curves were divided onto two groups, depending on the activity state of the system. Within each group the average detrended light curves were calculated  
(Fig. 2).

There is well pronounced secondary eclipse on a phase 0.5 on the light curve both in the outburst and the quiescent states. This eclipse occurs when 
the secondary (red dwarf) star is behind the white dwarf and the disc.
Our data shows that the amplitudes of the secondary minima in the quiescent and the outburst states are constant.
Thus, an occulting media has the same properties in both states because the secondary star has the constant luminosity expected.

The shape of quiescent state light curve is typical for the systems with significant ellipsoidality effect.

\begin{figure}[t]
\includegraphics[width=84mm]{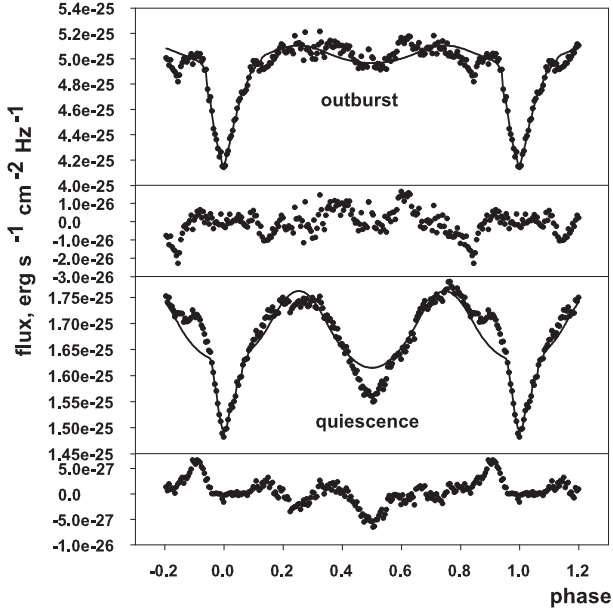}
\caption{Averaged light curves for two different states of the brightness of the system, the eclipse model fits and the residuals.}
\end{figure}

\section{Eclipse mapping of EM Cyg} 

Light curves of the partial eclipse of accretion disc in EM Cyg allow us to reconstruct the
temperature distribution in the eclipsed parts of accretion disc using the eclipse mapping technique \citep{hor85}. If we use the system 
parameters, determined by \citet{nor00} and \citet{wel07} ($i=67^{\circ}, M_{rd}=0.77, M_{wd}=1.0$), one can see that there is no eclipse 
of the white dwarf in the system (see Fig.3 in \citet{rob74}). The eclipse light curves become more sensitive to the accretion disc structure fluctuations in 
such situation. To minimize the influence of accretion disc non-uniformity, we used the averaged light curves for the outburst and the quiescent 
activity states of EM Cyg, presented in the Fig.~2.

\begin{table}[t]
 \centering
  \center
  \caption{The system parameters of EM Cyg, used in the eclipse models.}
  \begin{tabular}{ccc}
  \hline
Parameter&value&reference\\
  \hline
$M_{wd}$&$1.0 M_{\odot}$&\citet{wel07}\\
$M_{rd}$&$0.77 M_{\odot}$&\citet{wel07}\\
$i$&$67^{\circ}$&\citet{nor00}\\        
$p$&0.290909199 d&see text\\        
$A_{R}$&0.1 kpc$^{-1}$& see text\\        
$d$&200-500 pc&\\        
\hline         
\end{tabular}  
\end{table}   

Our model takes into account radiation of the eclipsed part of the accretion disc, the donor star light with ellipsoidality and gravitation darkening effects
and the constant term, wich includes the radiation from the uneclipsed part of accretion disc and the third star light. So, we cannot resolve 
the third star light fraction in the constant radiation parameter but the eclipse mapping results are independent 
from the presence of any additional constant sources.

Our eclipse mapping technique \citep{hal07} based on genetic algorithm fitting of the light curves with the model, where brightness distribution in two-dimensional
accretion disc is modeled by the set of points. The density of such points is proportional to the brightness of the selected area of accretion disc. 
The advantage of this method is the smaller number of free parameters of the model than in the classical eclipse mapping techniques are used.

To model of the donor star radiation we suggested that it fills the Roche lobe and use the gravity darkening parameter $\beta$ to be 1. 
The reflection effect was neglected because it is expected to be small in a late-type companion.

The maps of the eclipsing part of accretion disc for the quiescent and the outburst states 
have been calculated using the system parameters, presented in Table~2. 
To build the map of accretion disc we used the model 
with 200 radiating points.

The fits and the residuals one can see in Fig.~2. In the quiescent state we see well pronounced pre-eclipse hump, corresponded to the hot
spot in accretion disc, where ballistic flow coincides with the disc.

In the quiescent state, the secondary star contributes of about 64 per cent of the light to the total flux from the system and the third star in R band.

The secondary eclipse deep is about 5.3 per cent of the flux from the red dwarf and the width of the eclipse is about 15 per cent of the orbital period.
Taking into account low inclination angle, the origin of covering might be ballistic stream or extended above orbital plane gaseous structures.

For our models we used correction for interstellar extinction according \citet{fer75}  $E_{R-I}\simeq0.74 E_{B-V}$ and $A_{R}\simeq3.0 E_{R-I}$. 
In the case of $E_{B-V}=0.05$  we have $A_{R}$ to be about 0.1 per kiloparsec.

\begin{figure}[t]
\includegraphics[width=\columnwidth]{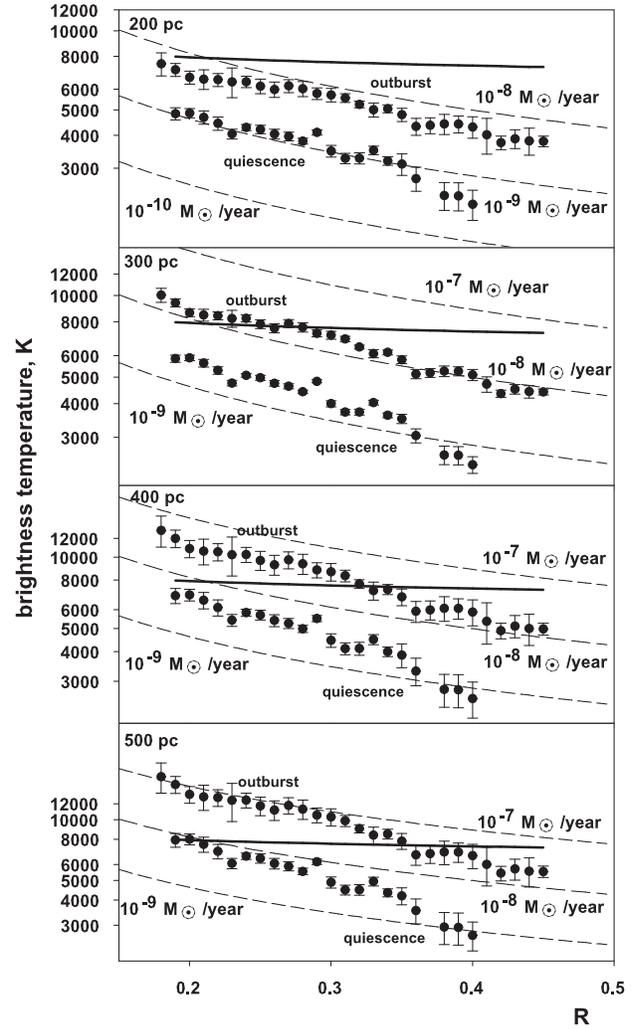}
\caption{Radial brightness temperature distribution in accretion disc for the outburst and the quiescent states for different distances to the system. 
The dashed lines are theoretical temperature distribution for the steady state flat disc and the thick solid line shows the critical temperature above 
which gas is in steady accretion regime \citep{war95}.}
\end{figure}

\begin{table}[h]
 \centering
  \center                                            
  \caption{Mass accretion rates ($M_{\odot}$/year) for different distances.}
  \begin{tabular}{c|cc}
  \hline
&\multicolumn{2}{c}{state}\\
distance, pc&quiescent&outburst\\
  \hline
200&$1.2\cdot 10^{-9}$&$5.6\cdot 10^{-9}$\\
300&$2.2\cdot 10^{-9}$&$1.4\cdot 10^{-8}$\\
400&$3.8\cdot 10^{-9}$&$3.1\cdot 10^{-8}$\\
500&$6.1\cdot 10^{-9}$&$6.2\cdot 10^{-8}$\\
\hline         
\end{tabular}  
\end{table}   

Taking into account of interstellar extinction, we calculated the brightness distribution in the eclipsing part of 
accretion disc for the distances of 200, 300, 400 and 500 parsecs. 

Using  the brightness distribution in accretion disc and the blackbody approximation, we calculated the brightness temperature distributions and compare it with predictions 
of the stationary accretion disc models. In Fig. 3 the radial brightness temperature plots for different distances are shown. 
The dashed lines are the temperature distribution predicted for the steady state solutions for different accretion rates. 
The thick solid line represents the critical temperature above which gas is ionized and supports the steady accretion regime \citep{war95}. 

When we use the distance to the system to be 200 pc, the temperature distribution fit is close to the steady state model distribution with 
$b = 3/4$ for the quiescent state and the outburst state both. At the same time, the critical temperature curve lies above the outburst 
temperature distribution curve.

In the case when the distance to be 500 pc the temperature distributions in the outburst and the quiescent states both are even steeper than
the steady state models what is uncommon.

From the temperature distribution curve in quiescent state we can find the size of accretion disc to be about $0.35a$ where $a$ is the binary separation.
This value slightly larger than that determined by \citet{go09} $0.3a$ from UV data.

Comparing the temperature distribution with the steady state models, we estimated the mass accretion rate for different distances to
the system in the quiescent and in the outburst states both (see Table 3). The mass accretion rate estimates were calculated by
minimizing of the mean square deviation of observed temperature values from the steady state temperature distribution for different accretion rates.

\section{Discussion} 

Our mass accretion rate estimates are close to values, determined by the previous investigators, if we choose the distance to be about 
200 pc. At the same time, the disc becomes ionized according to \citet{war95} critical temperature curve during outburst for the models with distance larger than 300 pc.
Distances about 500 pc demand ionized gas state in the disc even for quiescent states, what is improbable. So, using ionization criteria 
during outburst state, we must accept the distances, ranged between 300 and 400 pc.

Another estimate of the distance we can make using ellipsoidal modulation of the light curve in the quiescent state, where such effect is
mostly pronounced. Using mean flux of the secondary from our model, we find that it has $m_R\approx13.6$. If the secondary star having 
mass to be $0.77 M_{\odot}$ is a normal main sequence star, the spectral class must be K0V. In that case $m_V=5.9$ \citep{al73}, $V-R=0.64$ \citep{j66} and
the distance must be about 459 pc. At the same time for the system parameters we used, to fill its Roche lobe, the secondary must be slightly evolved.
If we use the spectral class value K2V, determined by \citet{jks81} the distance estimate will be 400 pc. This value is the same as determined by
\citet{ws03} using the $M_v$ versus $P_{orb}$ relation.

The distance of 400 pc in our models corresponds to relatively high accretion rate. Taking into account that the outburst state and the quiescent state
have practically equal durations
we estimate of the mass transfer rate to be about $1.7\cdot 10^{-8} M_{\odot}/year$. This value about an order higher than determined by \citet{cz08}
upper limit $2\cdot 10^{-9} M_{\odot}/year$ for conservative mass transfer. To explain of observed stability of the orbital period in the case of high mass
transfer rate we can take into account of magnetic braking effect. Using relations from \citet{cz08} we find that in the case of system parameters from
Table 2, the mass transfer rate which can be neutralized by magnetic braking is about $2.5\cdot 10^{-8} M_{\odot}/year$. So, magnetic braking of
EM Cyg system can explain of the mass transfer rate wich we have found for the case of 400 pc distance. Our present result supports the suggestion 
of \citet{cz08} that these two effects neutralize each other and that is why we do not see any observable period variation.
                                            
\section{Conclusions}
In this paper, using large volume of minima timings we specified the orbital period of the system. Our R band light curves in the quiescent state show
significant ellipsoidality effect. Firstly for this star we used the eclipse mapping technique of the outer parts of accretion disc. 
We estimated the mass accretion rate for different distances and showed, that the distance must be in the range between 300 and 400 pc. 
Using ellipsoidal variability of the secondary star we estimated the distance to be about 400 pc. Also we showed, that the magnetic braking effect
can explain of $1.7\cdot 10^{-8} M_{\odot}/year$ mass transfer rate, which has been found for 400 pc distance.

\acknowledgments

We thank to anonymous referee for many useful comments. Also we are grateful to I. Andronov, who proposed this star for investigations. 
We acknowledge with thanks to K. Antoniuk and A. Lomach for their help
during observations. We also thank to E. Sion and other colleagues for many fruitful discussions.


\begin{thebibliography}{}
\bibitem[Allen(1973)]{al73} Allen, C.W.  2004. Astrophysical Quantities, 3rd edn, Athlone press, London
\bibitem[Andronov \& Baklanov(2004)]{ab04} Andronov, I.L. \& Baklanov, A.V.  2004, Astron. School Rep., 5, 264
\bibitem[Andronov \& Marsakova(2006)]{am06} Andronov, I. L. \& Marsakova, V. I. 2006, JAAVSO, 35, 198
\bibitem[Bailey(1981)]{bay81} Bailey, J.  1981, \mnras, 197, 31
\bibitem[Bruch \& Engel(1994)]{be94} Bruch, A. \& Engel, A. 1994, \aaps, 104, 79
\bibitem[Beuermann \& Pakull(1984)]{bp84} Beuermann, K. \& Pakull, M. W. 1984, \aap, 136, 250
\bibitem[Brady \& Herczeg(1977)]{br77} Brady, R. A. \& Herczeg, T. J. 1977, \pasp, 89, 71
\bibitem[Csizmadia et al.(2008)]{cz08} Csizmadia, Sz., Nagy, Zs., Borkovits, T., Hegedus, T., Biro, I. B. \& Kiss, Z. T. 2008, Astronomishe Nachrichten, 329, 39
\bibitem[Fernie(1975)]{fer75} Fernie, 1975, Variable Stars And Stellar Evolution. Editor: L. Plaut. Kluwer Academic Publishers Group (Netherlands)
\bibitem[Godon et al.(2009)]{go09} Godon, P., Sion, E.M., Barrett, P.E. \& Szkody, P. 2009, \apj, 701, 1091
\bibitem[Halevin(2007)]{hal07} Halevin, A.V. 2007, Odessa Astronomy Publications, 20, 70
\bibitem[Horne(1985)]{hor85} Horne, K. 1985, \mnras, 213, 129
\bibitem[Hoffmeister(1928)]{hof28} Hoffmeister, C. 1928, Astronomishe Nachrichten, 233, 33
\bibitem[Hroch(1998)]{hro98} Hroch, F. 1998, Proceedings of the 29th Conference on Variable Star Research. 7th - 9th November 1997. Brno, Czech Republic. editor J. Dusek and M. Zejda., 30
\bibitem[Jameson et al.(1981)]{jks81} Jameson, R.F., King, A.R. \& Sherrington, M.R. 1981, \mnras, 195, 235
\bibitem[Johnson(1966)]{j66} Johnson, H.L.. 1966, A.Rev.Astr.Astrophys., 4, 193
\bibitem[Kim et al.(2005)]{kim05} Kim, Y.G., Andronov, I.L., Park, S.S. \& Jeon, Y.-B. 2005, \aap, 441, 663
\bibitem[La Dous(1991)]{lad91} La Dous, C. 1991, \aap, 252, 100
\bibitem[Misselt(1996)]{mis96} Misselt, K.A. 1996, \pasp, 108, 146
\bibitem[Mumford \& Krzeminski(1969)]{mk69} Mumford, G.S. \& Krzeminski, W. 1969, \apjs, 18, 429
\bibitem[Mumford(1974)]{mum74} Mumford, G.S. 1974, IBVS, 889, 1
\bibitem[Mumford(1975)]{mum75} Mumford, G.S. 1975, IBVS, 1043, 1
\bibitem[Mumford(1980)]{mum80} Mumford, G.S. 1980, \aj, 85, 748
\bibitem[North et al.(2000)]{nor00} North, R. C., Marsh, T. R., Moran, C. K. J., Kolb, U., Smith, R. C. \& Stehle, R. 2000, \mnras, 313, 383
\bibitem[Robinson(1974)]{rob74} Robinson, E. L. 1974, \apj, 193, 191
\bibitem[Stiening et al.(1982)]{sti82} Stiening, R. F., Dragovan, M. \& Hildebrand, R. H. 1982, \pasp, 94, 672
\bibitem[Stover et al.(1981)]{sto81} Stover, R. J., Robinson, E. L. \& Nather, R. E. 1981, \apj, 248, 696
\bibitem[Verbunt(1987)]{ver87} Verbunt, F. 1987, \aaps, 71, 339
\bibitem[Voikhanskaia et al.(1978)]{voi78} Voikhanskaia, N. F., Gnedin, Iu. N., Efimov, Iu. S., Mitrofanov, I. G. \& Shakhovskoi, N. M. 1978, Soviet Astronomy Letters, 4, 148
\bibitem[Warner(1995)]{war95} Warner, B.  1995, Cambr. Astrophys. Ser. 28, "Cataclysmic Variable Stars". Cambridge Univ. Press, Cambridge 
\bibitem[Welsh et al.(2005)]{wel05} Welsh, W.F., Froning, C.S., Marsh, T.R., Robinson, E.L. \& Wood, P.R. 2005, \pasp, 330, 351
\bibitem[Welsh et al.(2007)]{wel07} Welsh, W.F., Froning, C.S., Marsh, T.R., Reimer, T.W., Robinson, E.L. \& Wood, P.R. 2007, \pasp, 362, 241
\bibitem[Winter \& Sion(2003)]{ws03} Winter, L. \& Sion, E. 2003, \apj, 582, 352
\end{thebibliography}

\end{document}